\documentclass[12pt,preprint]{aastex}
\usepackage{emulateapj5,epsf}

\usepackage{lscape}
\makeatletter
\newcounter{subeqncnt}
\def\thesubeqncnt{\alph{subeqncnt}}
\def\subequations{\begingroup%
\stepcounter{equation}\edef\@tempa{\theequation}%
\let\c@equation\c@subeqncnt\c@subeqncnt\z@
\edef\theequation{\@tempa\noexpand\thesubeqncnt}}

\makeatother

\shorttitle{
Fate of Young Radio Galaxies
}
\shortauthors{Kawakatu, Nagai and Kino}

\begin{document}

\title{
The Fate of Young Radio Galaxies: 
Decelerations Inside Host Galaxies?
}


\author{Nozomu Kawakatu\altaffilmark{1}}
\affil{National Astronomical Observatory of Japan, 2-21-1 Osawa, 
Mitaka, Tokyo 181-8588, Japan}

\author{Hiroshi Nagai}
\affil{National Astronomical Observatory of Japan, 2-21-1, 
Osawa, Mitaka, Tokyo 181-8588, Japan}

\author{Motoki Kino}
\affil{Institute of Space Astronautical Science,
JAXA, 3-1-1 Yoshinodai, Sagamihara 229-8510, Japan}


\altaffiltext{1}{kawakatu@th.nao.ac.jp}

\begin{abstract}
We examine the evolution of variously-sized radio galaxies [i.e., compact symmetric objects (CSOs), medium-size symmetric objects (MSOs), Fanaroff-Riley type 
I\hspace{-.1em}I radio galaxies (FRI\hspace{-.1em}Is)], by comparing 
the relation between the hot spot size and the projected linear 
size with a coevolution model of hot spots and a cocoon. 
We take account of the deceleration effect by the cocoon head growth. 
We find  that the advance speed of hot spots and lobes 
inevitably show the deceleration phase (CSO-MSO phase) and the 
acceleration phase (MSO-FRI\hspace{-.1em}I phase). This is ascribed to the 
change of the power-law index of ambient density profile in the MSO phase (` 1 kpc). It is also found that the cocoon shape becomes nearly spherical 
or disrupted for MSOs, while an elongated morphology is predicted for CSOs 
and FRI\hspace{-.1em}Is. 
This seems to be consistent with the higher fraction of distorted morphology 
of MSOs than that of CSOs and FRI\hspace{-.1em}Is. Finally, we predict that 
only CSOs whose initial advance speed is higher than about 0.1c can evolve 
into FRIIs, comparing the hot spot speed with the sound speed of 
the ambient medium. 

\end{abstract}

\section{Introduction} 
According to the standard model for large-scale radio sources 
(Blandford \& Rees 1974), a pair of relativistic jets transport away 
the bulk kinetic energy from the central compact 
region close to a central supermassive black hole 
(SMBH) to $\sim 100\,{\rm kpc}-10^{3}\, {\rm kpc}$ scale radio lobes. 
This bulk kinetic energy is dissipated by the strong terminal shocks 
which are identified as hot spots. 
Then, shocked relativistic plasma expands sideways and envelopes 
the whole jet system and this is a so-called cocoon (Scheuer 1974). 
Now, how young AGN jets (or radio sources) evolve into 
the extended radio sources (e.g., FRI\hspace{-.1em}I radio galaxies) 
is one of the primal problems in astrophysics. 
In order to clarify this issue, it is important to understand the 
nature of small, young progenitors of FR I\hspace{-.1em}Is. 
The study of young progenitor of FRI\hspace{-.1em}Is 
is also a good opportunity to get information about the central regions 
of galaxies, especially their mass density distribution. 
The coevolution of SMBHs and galaxies has been discussed 
by many authors who say that AGN feedback may be a key mechanism in 
regulating the growth of SMBHs and also in preventing further star formation 
of galaxies; These authors use an analytical approach (e.g., Silk \& Rees 1998; King 2003; Wyithe \& Loeb 2003; Murray et al. 2005) and 
a semi-analytical approach (e.g., Granato et al. 2004; Kawata \& Gibson 2005; 
Di Matteo et al. 2005; Okamoto et al. 2007). 
However, it has been unclear how we should treat AGN feedback associated with 
AGN jets, because we do not understand how small, young radio sources 
evolve into FRI\hspace{-.1em}Is. 
Therefore, for understanding AGN feedback, it is essential to explore 
conditions under which small, young radio sources can pass through a dense 
ambient medium within host galaxies.

Recently, a large number of compact, bright double-lobe radio sources, 
so-called ``compact symmetric objects (CSOs)'' whose linear size 
is $\leq 1\,{\rm kpc}$ have been discovered 
(Wilkinson et al. 1994; Fanti et al. 1995; Readhead et al. 1996a, b). 
Concerning the origin of CSOs, two scenarios were proposed. 
One is a so-called ``frustrated jet scenario'' in which the ambient medium 
is so dense that the jet cannot break through, so that 
sources are old and confined (van Breugel, Miley \& Heckman 1984). 
The other is  a ``youth radio source scenario'' in which CSOs are the young 
progenitor of FRI\hspace{-.1em}I radio galaxies (e.g., Philips \& Mutel 1982).
The former scenario looks less likely because recent multi-wavelength studies 
show that the ambient medium of CSOs is not significantly different from 
FRI\hspace{-.1em}I radio galaxies (O'Dea et al. 1998 and references therein). 
On the other hand, the most compelling evidence for the youth jet scenario 
is found in the recent observations of advance speed of hot spots 
of several CSOs (Owaiank, Conway \& Polatidis 1998; 
Taylor et al. 2000; Tschager et al. 2000; Giroletti et al. 2003; 
Polatidis \& Conway 2003; Gugliucci et al. 2005; 
Nagai et al. 2006; Gugliucci et al. 2007; Luo et al. 2007). 
They showed that the separation velocity is typically $\sim 0.1\, c$, 
indicating a dynamical age of $\sim 10^{2-4}\,{\rm yr}$ 
which implies that CSOs are possible candidates as the progenitors 
of FRI\hspace{-.1em}Is.

Based on the youth scenario, the evolution of radio sources has been considered by a number of authors (Carvalho 1985; Begelman \& Cioffi 1989, hereafter BC89; Falle 1991; Fanti et al. 1995; Readhead et al. 1996a; Begelman 1996; De Young 1997; Kaiser \& Alexander 1997; Alexander 2000; Snellen et al. 2000; Perucho \& Mart{\'{\i}} 2002; Kawakatu \& Kino 2006, 
hereafter KK06). 
For simplicity, one has supposed a constant advance speed of hot spots, 
or a constant aspect ratio of the cocoon (i.e., a self-similar evolution) for 
a long-term evolution of radio-loud AGNs. 
{\it Do these treatments reflect the actual evolution of radio sources ?} 
The answer is not trivial.
According to previous observations, a stronger interaction with the ambient 
medium has been reported for more compact radio sources 
(e.g., Gelderman \& Whittle 1994; de Vries et al. 1997; Axon et al. 2000; 
Gupta et al. 2005; Holt et al. 2008). Thus, the advance speed of hot spots 
might decelerate via the strong interaction in host galaxies. 
On the evolution of cocoon morphology, recent observations have shown 
that the fraction of distorted radio morphology of medium-size 
symmetric objects (MSOs) is higher than CSOs and FRI\hspace{-.1em}Is 
(e.g., Saikia et al. 1995; O'Dea 1998; Dallacasa et al. 2002a, b). 
Observationally, it is unclear whether the assumptions 
(constant hot spot velocity or constant aspect ratio) 
are reasonable. 
Also, the validity of these assumptions is still under debate theoretically
(De Young 1997; Komissarov \& Falle 1998; O'Dea 1998; Carvalho \& 
O'Dea 2002a, b; Perucho \& Mart{\'{\i}} 2003: hereafter PM03). 
Therefore, in order to answer the above question, it is essential to build up 
an appropriate model of radio sources {\it without assuming the 
constant aspect ratio and the constant advance speed of hot spots} presented 
by KK06.

Recently, it has been found that the power law index for the evolution 
of hot spot size changes at the transition between the interstellar medium 
and the intergalactic medium, i.e., $\sim 1-10\,{\rm kpc}$ 
(Jeyakumar \& Saikia 2000: hereafter J00; PM03). 
Since the hot spot is one of the most important ingredients 
in the whole jet system, the evolution of hot spot size would reflect 
the dynamical growth of radio sources of various scales. 
However, it was difficult to derive the dynamical evolution of radio 
sources from previous work (J00 and PM03) because of the lack of spatial 
resolution, the observational bias and small sample of radio sources. 
Therefore, we should not only compile a larger sample of CSOs, MSOs and 
FRI\hspace{-.1em}Is, but also be careful about the data quality and the 
observational bias. 
Then, by comparing the direct comparison with KK06, 
we examine the deceleration and acceleration of hot spots.

The paper is organized as follows: 
In $\S2$, we first elucidate whether the power-law index 
change at $\sim$ 1 kpc is a solid result for the evolution of hot spot size 
($\S2.1$).
To this end, we compile  a larger sample of CSOs, MSOs and FRI\hspace{-.1em}I 
sources than previous work (J00; PM03) by considering the observational bias 
and being careful about the data quality. 
We also discuss the advance speed of the hot spot ($\S2.2$). 
In $\S 3$, to interpret these observational data, 
we briefly review the physical model of hot spot evolution (KK06) 
with the aid of cocoon dynamics (Kino \& Kawakatu 2005; hereafter KK05). 
In $\S 4$, from the direct comparison with KK06 
we reveal the deceleration and acceleration of advance speed of hot spots, 
and the evolution of cocoon morphology. 
In $\S 5$, we predict the fate of CSOs by comparing the evolution of hot spot 
velocity with the sound speed of the ambient medium. 
Finally, we discuss the FRI/I\hspace{-.1em}I dichotomy and AGN feedback 
due to the cocoon (i.e., AGN bubble) expansion. 
Section 6 is devoted to our summary. 
Throughout this paper, we adopt $\mathrm H_{0}=72$~km s$^{-1}$Mpc$^{-1}$ 
and $\mathrm q_{0}=0.5$.

\section{Observational properties of hot spots}
Following the recent work of Nagai (2007), we compiled variously-sized radio galaxies and examined physical quantities of the hot spot (Table 1).  FRI\hspace{-.1em}I sources ($l_{\rm h}\gtrsim10$~kpc, where $l_{\rm h}$ is linear size of radio source) in our sample are selected from FRI\hspace{-.1em}I radio galaxies in the 3CR catalogue (Hardcastle et al. 1998; Fernini et al. 1993; Bridle et al. 1994; Gilbert et al. 2004; Kharb et al. 2008).  MSOs ($1\lesssim l_{\rm h} \lesssim10$~kpc) are selected from the samples in Dallacasa et al. (2002a) and Dallacasa et al. (2002b).  We also included Comact-Steep Spectrum (CSS) sources having double-lobe-like structure in Fanti et al. (1985) and Sanghera et al. (1995) in our MSOs sample.  CSOs ($\lesssim1$~kpc) are selected from the samples in Readhead et al. (1996b), Taylor et al. (2000), Wang et al. (2003), Polatidis et al. (2003), Maness et al. (2004) and Nagai et al. (2006).  In addition to these data, we measured hot spot parameters of a CSO (B~1943+546) by analyzing the data obtained from the NASA/IPAC Extragalactic Database (NED).  Sources with a complex and highly distorted structure have been excluded from the above samples.  To avoid the difference in estimation of physical quantities between the hot spot and the counter hot spot due to the projection effect, we focus on sources with relatively symmetric radio lobes in this work.  Finally, we compiled a total of 117 radio sources.  For the sources with multiple hot spots, we followed the method of hot-spot identification described in the original papers.  If there was no information of identification method available in the original papers, we referred to the most prominent component as the hot spot.  Based on these data, we investigate how the size of the hot spot changes with the distance from the core in $\S 2.1$.  We also discuss the advance speed of the hot spot in $\S 2.2$.

\vspace{5mm}
\epsfxsize=8cm 
\epsfbox{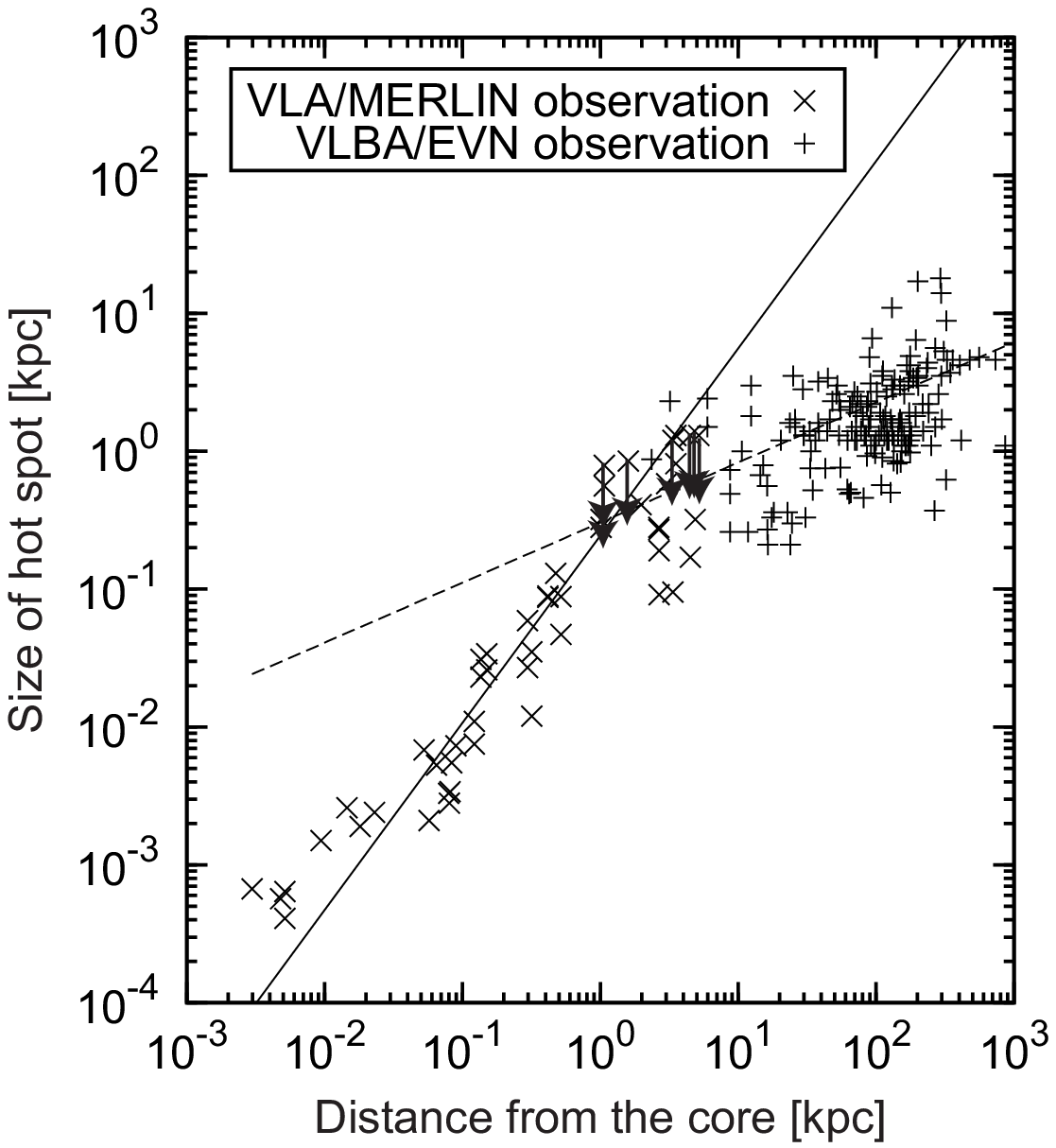}
\figcaption
{
The relation of hot-spot size ($r_{\rm HS}$) and 
hot-spot distance from the core ($l_{\rm h}$).  
Crosses with arrows indicate upper limit.  
Solid line corresponds to the best-fit for the sources 
where $10^{-3}$~kpc$\le l_{\rm h}\le1$~kpc whereas broken line corresponds 
to that for the sources where $1$~kpc$\le l_{\rm h}\le10^{3}$~kpc. 
Note that upper-limit data were not included in the fittings.
}
\vspace{2mm}

\subsection{Size}
Figure 1 shows the size of a hot spot, $r_{\rm HS}$ with respect to $\l_{\rm h}$, together with the best linear fit on the log-log plane ($\log{y}=a\log{x}+b$).  We estimated $r_{\rm HS}$ as $(\theta_{\rm maj}+\theta_{\rm min})/2$ where $\theta_{\rm maj}$ and $\theta_{\rm min}$ are full width at half maximum (FWHM) of major and minor axes of the component obtained by Gaussian model fittings, respectively.  The slope is changed around $\sim1$~kpc, and then the best fit value of $a$ for $l_{\rm h} < 1\,{\rm kpc}$ is $a=1.34\pm0.24$, while $a=0.44\pm 0.08$ for $l_{\rm h} > 1\,{\rm kpc}$.  This tendency is almost consistent with PM03, but a large number of sources in our sample allow us to confirm the tendency more clearly.  One could think the apparent slope change is caused by observational bias.  Here we consider the possible uncertainties of the estimation of the hot-spot size and the slope change.

Firstly, the sizes of hot spots of larger sources could be underestimated due to a lack of sensitivity for diffuse emission.  In this context, there are two major candidates for underestimation of size.  One is the missing flux for the diffuse emission due to the absence of short baseline of the interferometer.  Generally, this effect is more serious in VLBI (e.g., VLBA, EVN) observation than in interferometer (e.g., VLA) observation.  The hot-spot sizes of most sources with $l_{\rm h}>1$~kpc are estimated using data taken with VLA while that with $l_{\rm h}<1$~kpc are estimated using data taken with VLBI.  For the sources with $l_{\rm h}>1$~kpc, the missing flux is not significant because almost the total extent of the radio lobe has been, in fact, detected in most sources in our sample.  The other candidate is the image dynamic range.  Poor dynamic range could contaminate the diffuse emission so that the hot-spot sizes could be underestimated.  The image dynamic range strongly depends on the distribution of sampled visibilities on the uv-coverage if we assume each visibility has no significant errors (Perley 1999).  Denser and more uniform sampling of visibilities on the uv-coverage increases the image dynamic range.  Thanks to dense and uniform sampling of the visibility, the image dynamic range of VLA observation attains up to several thousands (e.g., Hardcastle et al. 1997), which is much higher than typical VLBI observations.  Therefore, we can safely exclude the possibilities of underestimation of the hot-spot sizes for larger sources.

Secondly, the angular resolution relative to the largest angular size of radio source ($n_{\rm b}$) possibly affects the estimation of the hot-spot size.  With high $n_{\rm b}$ value one could expect to see smaller-scale structures within the hot spots, which leads to a decrease in the sizes of hot spots.  FRI\hspace{-.1em}I sources in our sample have observed with $n_{\rm b}$ that is a few tens of times higher than $n_{\rm b}$ of CSOs and MSOs in our sample.  In this case, smaller scale structure within hot spots could be seen for FRI\hspace{-.1em}I sources than those of CSOs and MSOs.  However, if CSOs and MSOs are observed with the same order of $n_{\rm b}$ for FRI\hspace{-.1em}I sources, the hot-spot sizes of CSOs and MSOs should decrease.  This effect is more serious in smaller sources.  As a result, the slope in CSOs-MSOs phase would be more inverted ($a>1.34\pm0.02$) but the starting point of slope change would be unchanged.  Intensive study by JS00 found a similar change in slope, but the change occurred around $\sim20$~kpc.  They selected the sample observed with similar $n_{\rm b}$ among CSOs, MSOs, and FRI\hspace{-.1em}I sources.  However, their sample contains many FRI\hspace{-.1em}I sources where their hot-spot sizes have been determined as only upper limits due to lack of angular resolution.  This would make the starting point of the slope change unclear.

Lastly, we comment on the projection effect.  To be precise, our estimated size of the hot spot is not intrinsic size but the size projected onto the celestial plane.  However, our sample was restricted to the sources with relatively symmetric lobes.  We excluded the one-sided sources.  Thus the difference in the jet axis is not so large among the sources, and affects the estimation of the size by no more than a factor of a few.  

Overall, even if we allow for possible uncertainties discussed above, the trend of the $r_{\rm HS}-l_{\rm h}$ relation would not be changed.  It seems reasonable to suppose that the slope change occurs around 1~kpc.

\subsection{Advance speed}
Figure 2 shows the advance speed of hot spots ($v_{\rm HS}$).  For CSOs, $v_{\rm HS}$ has been directly measured by VLBI observations.  In the case of CSOs where the hot spot motion was measured relative to the counter hot spot (so-called the separation rate: $v_{\rm sep}$), we estimated $v_{\rm HS}$ as $v_{\rm sep}/2$.  On the other hand, $v_{\rm HS}$ of FRI\hspace{-.1em}I sources has not been measured directly.  We therefore adopted $v_{\rm HS}$ of MSOs and FRI\hspace{-.1em}I sources estimated by the synchrotron age constraints (Alexander et al.1984; Carilli et al.1991; Liu et al.1992; Klein et al. 1995; Mack et al. 1998; Parma et al. 1999; Murgia et al. 1998; Schoenmakers et al. 2000; Jamrozy et al. 2005), which is estimated by the break frequency of radio spectra caused by the synchrotron aging.  We adopted $v_{\rm HS}$ of MSOs only from the lobe-dominated source (classified gtype ah in Murgia et al. 1999).  We summarize the data of $v_{\rm HS}$ in Table 2.  

\vspace{5mm}
\epsfxsize=8cm 
\epsfbox{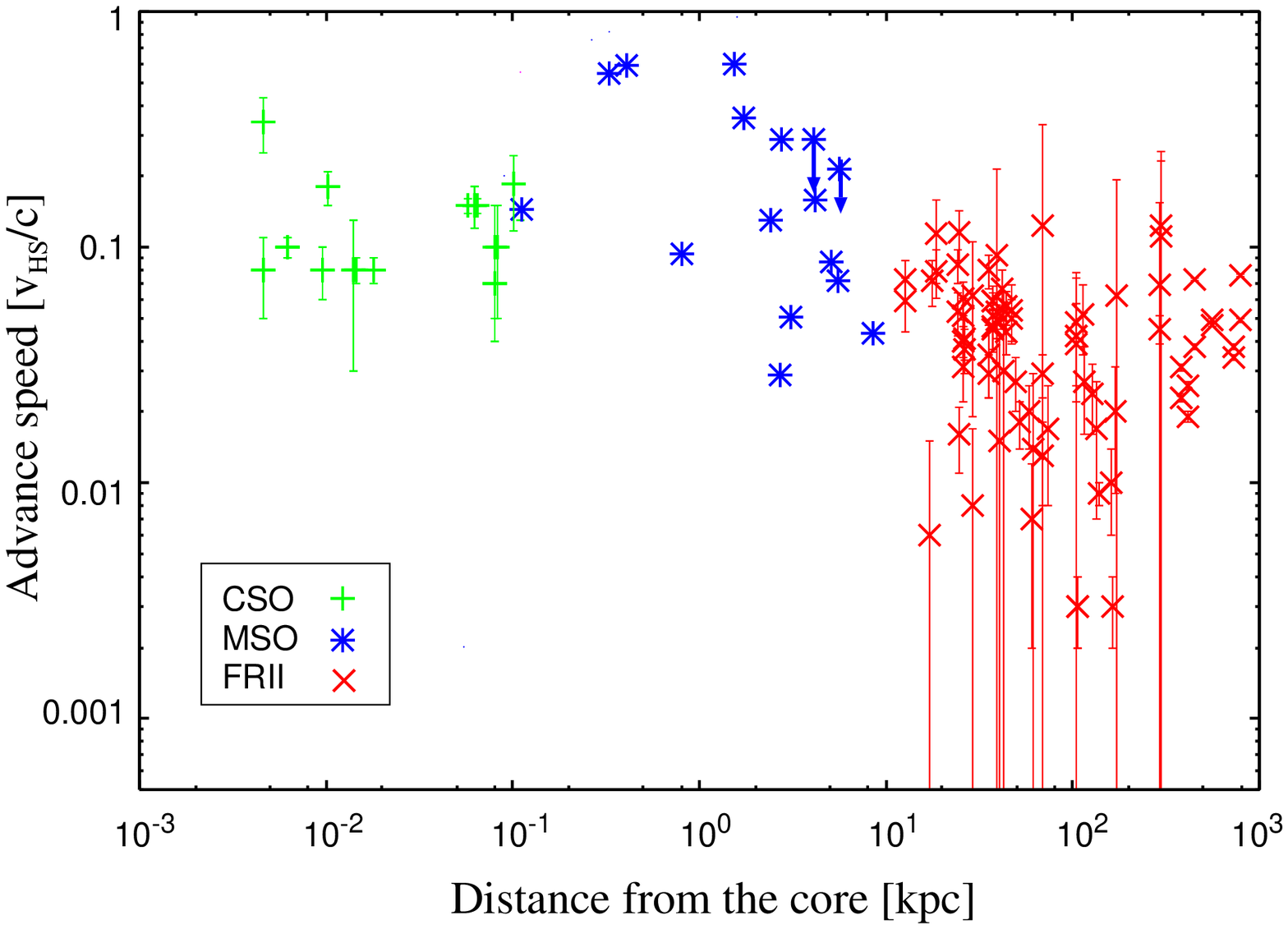}
\figcaption
{
The relation of hot-spot velocity ($v_{\rm HS}$) 
and hot-spot distance from the core ($l_{\rm h}$). 
Pluses (green), stars (blue) and crosses (red) represent 
CSOs, MSOs and FRI\hspace{-.1em}Is, respectively. 
}
\vspace{2mm}

As we will show in the followings, there are a number of biases and uncertainties in the estimation of $v_{\rm HS}$.  We should conclude that the evolution of $v_{\rm HS}$ relative to $l_{\rm h}$ is not clear at this moment.  We do not therefore argue the favor of the evolutional model based on the observational data of $v_{\rm HS}$.

Firstly, the apparent absence of CSOs with $v_{\rm HS}<\sim 0.1c$ is possibly due to the observational biases.  In Figure 3 we show the apparent velocity of the hot spot of CSOs with respect to redshift of the source.  Each parameter is listed in Table 3.  Apparent velocity of less than $\sim10$~$\mu$as/yr has not been detected.  This is probably due to a lack of angular resolution.  Apparent velocity of $\sim10$~$\mu$as/yr corresponds to $\sim0.1c$ in $z>0.1$, so that it is possible to conclude that the apparent absence of $v_{\rm HS}<0.1c$ in CSOs is due to biases introduced by the detection limits.  

Secondly, the synchrotron age estimates are more uncertain compared to the kinematic age estimates.  Primary uncertainty in the estimation of synchrotron age is the magnetic field.  Although the minimum-energy condition is highly uncertain, the minimum-energy field is usually adopted due to the lack of better estimation.  Since the synchrotron age strongly depends on the magnetic field rather than the break frequency ($t_{\rm syn}\propto B^{-1.5}\nu_{\rm b}^{-0.5}$), uncertainty of magnetic field makes it difficult to relate the source age.  Therefore, $v_{\rm HS}$ of MSOs and FRI\hspace{-.1em}I sources are more unreliable than that of CSOs.

\vspace{5mm}
\epsfxsize=8cm 
\epsfbox{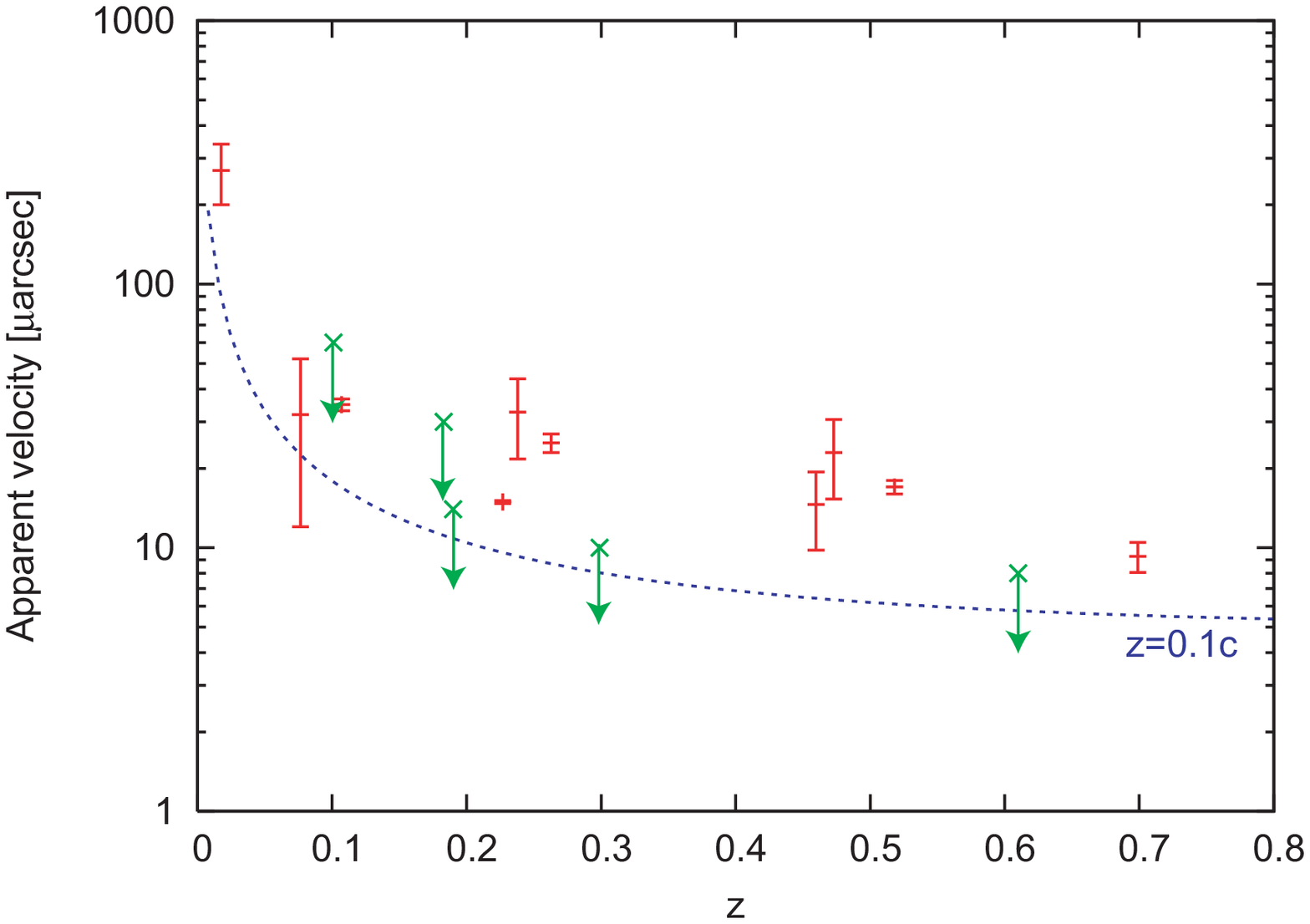}
\figcaption
{
Apparent separation velocity between hot spots of CSOs.  Arrows indicate 
upper limits (Taylor \& Vermeulen 1997; Gugliucci et al. 2005). 
Concerning the sources in Gugliucci et al. (2005), we have not adopted J~1415+13 because separation rate is measured not between hot spots but jet components.  Broken line corresponds to the velocity of $0.1c$.
}
\vspace{2mm}

Finally, a direct comparison of $v_{\rm HS}$ between MSOs and FRI\hspace{-.1em}I sources is not acceptable because the estimation method of $v_{\rm HS}$ is different between MSOs and FRI\hspace{-.1em}I sources.  Basically, $v_{\rm HS}$ of FRI\hspace{-.1em}I sources is calculated by the linear fit to the observed synchrotron age distribution across the lobe as $v_{\rm HS}$ of FRI\hspace{-.1em}I sources (cf. Fig. 3 in Alexander et al. 1984).  On the other hand, the advance velocities of MSOs in Murgia et al. (1999) are estimated in a different way.  Murgia et al. calculated the break frequency by fitting the integrated spectra to the continuous injection model (CI model).  Then they estimated $v_{\rm HS}$ as $l_{\rm h}/t_{\rm CI}$ where $t_{\rm CI}$ is the synchrotron age estimated from the break frequency of the CI model spectrum.  The main concern of this method is that the integrated spectrum could contain not only the lobe emission but also the core and jet emission.  To be precise, the estimated synchrotron age is the gmeanh particle age of these components.  The electrons in the core and jets, especially in the knots of the jets, are likely to be sites of particle acceleration.  Thus the estimated synchrotron age could be underestimated compared to actual source age, which leads to the overestimation of $v_{\rm HS}$.  

\section{A coevolution model of hot spots and a cocoon} 
In this section, we briefly review a dynamical evolution model of 
radio sources (KK06) which traces the dynamical evolution of 
advancing hot spots  and expanding cocoon (e.g., BC89; see also KK05). 
Hereafter, the mass, momentum and kinetic energy of unshocked jet flow 
are conserved. In other words, we do not consider the entrainment effect 
of the ambient medium. This would be justified for the relativistic jet flows 
(e.g., Scheck et al. 2002: hereafter S02; Mizuta, Yamada \& Takabe 2004).

The basic equations of the cocoon expansion can be obtained as follows:

\begin{enumerate}
\item
The equation of motion along the jet axis, i.e., 
the momentum flux of a relativistic jet is balanced to the ram 
pressure of the ambient medium spread over the effective cross-section 
area of the cocoon head, $A_{\rm h}$, 
\begin{eqnarray}
L_{\rm j}/c&=&\rho_{\rm a}v_{\rm HS}^{2}A_{\rm h}, \nonumber
\end{eqnarray}
where $L_{\rm j}$, $\rho_{\rm a}$ and $v_{\rm HS}$ are the 
total kinetic energy of jets, the mass density of the ambient medium 
and the hot spot velocity, respectively. 
Here we assume that $L_{\rm j}$ is constant in time, and 
that $v_{\rm HS}=v_{\rm h}$ where $v_{\rm h}$ is the advance 
speed of the cocoon head (see KK06 for details).

\item
The equation of motion perpendicular to the jet axis, 
that is, the sideways expansion velocity, $v_{\rm c}$ is equal to 
the shock speed driven by the overpressured cocoon with the internal 
pressure, $P_{\rm c}$, 
\begin{eqnarray}
P_{\rm c}&=&\rho_{\rm a}v_{\rm c}^{2}. \nonumber
\end{eqnarray}
For CSOs, the sideways expansion velocity evaluated by the outflow 
velocity of the line-emitting gas (e.g., de Veries et al. 1999; Axon et al. 
2000; O'Dea et al. 2002; Gupta et al. 2005; Holt et al. 2008) would be higher 
than the sound velocity of the ambient medium. 
Even for FRI\hspace{-.1em}Is, the overpressured expansion would be valid 
(e.g., KK05; Ito et al. 2008).

\item 
The energy conservation in the cocoon, namely 
all of the kinetic energy transported by jets during the source age $t$ 
is is deposited as the cocoon's internal pressure, 
\begin{eqnarray}
P_{\rm c}V_{\rm c}&=&2(\gamma_{\rm c}-1)L_{\rm j}t, \nonumber
\end{eqnarray}
where $V_{\rm c}$ is the the volume of the cocoon and $\gamma_{\rm c}=4/3$ 
is the specific heat ratio of the relativistic plasma in the cocoon. 
For simplicity, we do not consider the Bremsstrahlung cooling of 
the cocoon, although this would be important in the much earlier phase 
of the evolution (Kino et al. 2007). 
\end{enumerate}

\vspace{5mm}
\epsfxsize=8cm 
\epsfbox{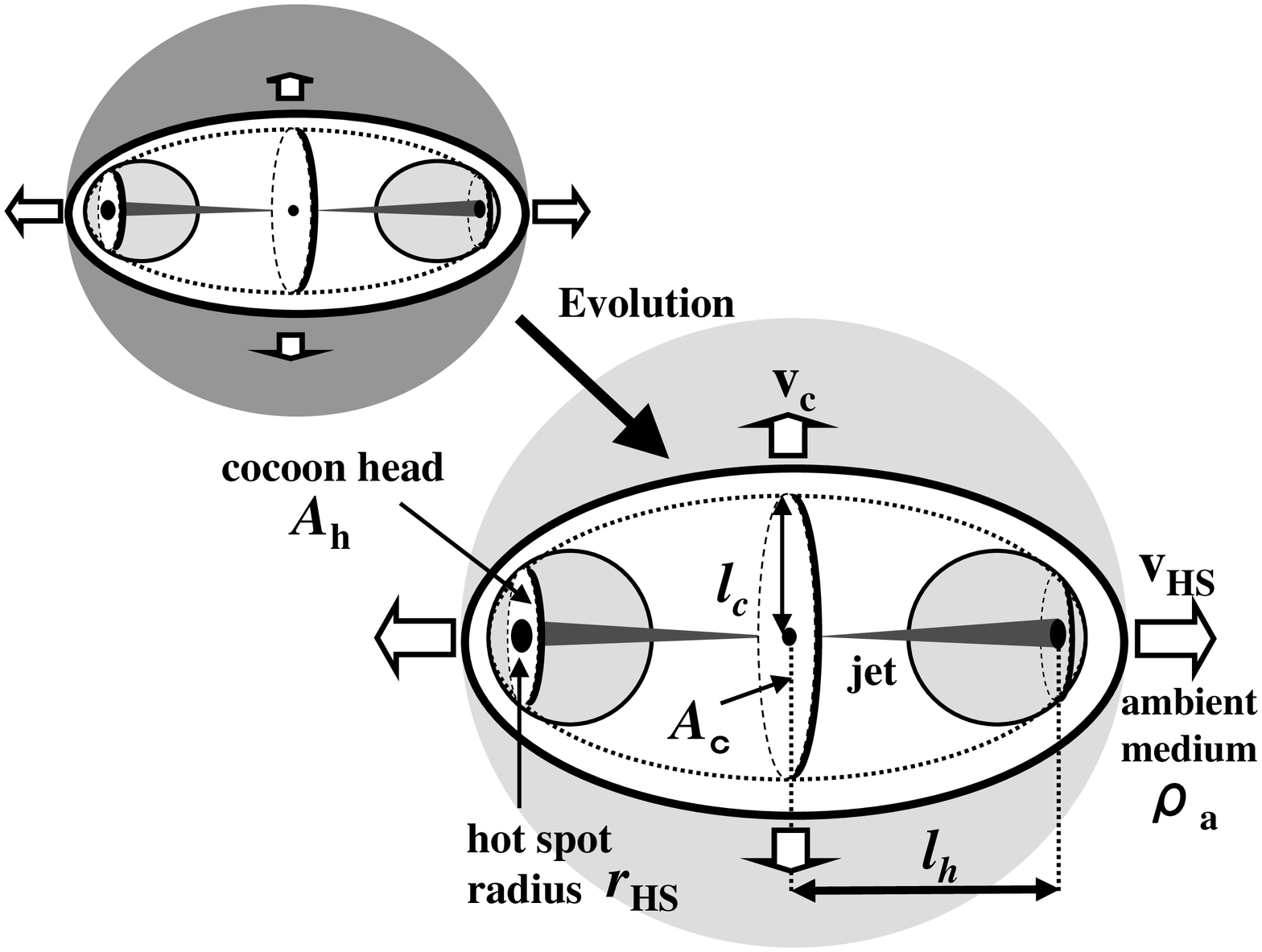}
\figcaption
{
A schematic picture of the evolution of radio sources. 
The aspect ratio of the cocoon is defined as ${\cal R}=l_{\rm c}/l_{\rm h}$.
}
\vspace{2mm}

According to KK06, the physical quantities of hot spots and the cocoon 
can be determined by the mass density of the ambient medium, 
$\rho_{\rm a}$ and the cross-section area of the cocoon body, 
$A_{\rm c}$ (see Figure 4). 
The mass density profile is expressed as 
\begin{eqnarray}
\rho_{\rm a}(d)&\propto& d^{-\alpha}, \nonumber
\end{eqnarray}
where $d$ is the radial distance from the center of the galaxy. 
The cross-sectional area of the cocoon body is given by 
\begin{eqnarray}
A_{\rm c}(t)&=&\pi l_{\rm c}^{2}\propto t^{X}, \nonumber
\end{eqnarray}
where $l_{\rm c}=\int_{t_{\rm min}}^{t} v_{\rm c}(t') d t'$ 
is the radius of the cocoon body. 
Here $t_{\rm min}$ is the time from when the two-dimensional (2D) phase 
($A_{\rm h}$ growth phase) starts.
On a free parameter $X$, we can constrain the value of X, 
in comparison with numerical simulations. 
We will discuss this issue later. 
Assuming that $A_{\rm h}/r_{\rm HS}^{2}$ is constant in time, 
the quantities of hot spots ($r_{\rm HS}$, $v_{\rm HS}$ and $P_{\rm HS}$) 
can be described in terms of the length from the center of the 
galaxy to the hot spot ($l_{\rm h}$) as follows:

\begin{eqnarray}
r_{\rm HS}&\propto&l_{\rm h}^{\frac{X(-2+0.5\alpha)(\alpha-2)+3\alpha-4}
{2X(-2+0.5\alpha)+6}}\propto {l_{\rm h}}^{S_{\rm r}}, \\
v_{\rm HS}&\propto& l_{\rm h}
^{\frac{2-X(2-0.5\alpha)}{X(-2+0.5\alpha)+3}} 
\propto {l_{\rm h}}^{S_{\rm v}}, \\
P_{\rm HS}&\propto&l_{\rm h}
^{\frac{X(2-0.5\alpha)
(\alpha-2)+4-3\alpha}{X(-2+0.5\alpha)+3}}
\propto {l_{\rm h}}^{S_{\rm p}},
\end{eqnarray}
where $S_{\rm r}$, $S_{\rm v}$ and $S_{\rm p}$ represent the values of 
the exponents of the hot spot radius, advance velocity and pressure, 
respectively. 
We apply the conditions of $Y\equiv X(-2+0.5\alpha)+3>0$, 
which make the contribution at $t_{\rm min}$ in the integration of 
$l_{\rm h}=\int_{t_{\rm min}}^{t} v_{\rm HS}(t') d t'\propto t^{Y}
-t_{\rm min}^{Y}$ small enough. In other words, we do not consider the case 
that the life-times of compact radio sources are much longer than those of 
extended radio sources because this is unrealistic. 

Here, let us consider the $\alpha$ and $X$ dependence on the advance 
velocity of hot spots for fixed physical quantities 
at $t=t_{\rm min}$. 
Concerning $\alpha$ dependence in fixed $X$, larger $\alpha$ leads to 
a weaker deceleration effect on $v_{\rm HS}$ due to smaller $\rho_{\rm a}$. 
Also, in order to keep the constant velocity of sideways expansion, 
larger $\alpha$ predicts smaller cocoon pressure $P_{\rm c}$ 
because of $P_{\rm c}=\rho_{\rm a}v_{\rm c}^{2}$. 
Next we consider the $X$ dependence. 
Larger $X$ leads to larger $A_{\rm c}$ and $v_{\rm c}$, and thus 
the larger $P_{\rm c}$ is required 
because of $P_{\rm c}=\rho_{\rm a}v_{\rm c}^{2}$. 
Thus, the slower $v_{\rm HS}$ is needed 
in order to satisfy the energy conservation in the cocoon, that is, 
$P_{\rm c}A_{\rm c}v_{\rm HS}=const.$ 
In summary, larger (smaller) $\alpha$ and smaller (larger) $X$ leads 
to acceleration (deceleration) of hot spot velocity.

The aspect ratio of the cocoon, ${\cal R}\equiv l_{\rm c}/l_{\rm h}$, 
is also an important quantity 
relevant to the evolution of radio galaxies.
The $l_{\rm h}$-dependence of the aspect ratio of cocoon is 
then given by 
\begin{equation}
{\cal R}\propto l_{\rm h}
^{\frac{X(2.5-0.5\alpha)-3}{X(-2+0.5\alpha)+3}} 
\propto l_{\rm h}^{S_{\rm R}}. 
\end{equation}

Finally, in order to check the reliability of their analytical model, 
we compare the results of KK06 compared with 2D relativistic hydrodynamic 
simulations in a uniform ambient medium with $\alpha=0$ (S02) and 
in a declining ambient medium with $\alpha=1$ 
(Perucho \& Marti 2007; hereafter PM07). 
We summarize the comparisons with 2D hydrodynamic simulations in Table 4. 
These results show that the analytic model (KK06) can well agree with 
the results of 2D simulations. 
Thus, it is possible to constrain the free parameter X 
by direct comparison with numerical simulations. 
Following the results of S02 and PM07, we can constrain $1.2 \leq X \leq 1.4$ 
for the wide range of $\alpha \,(=0-2.5)$ because $X$ does not depend on 
$\alpha$ significantly (see Table 4). 
As seen in Table 4 we find that the power law index of $r_{\rm HS}$ is 
slightly different from the results of KK06 and those of numerical 
simulations (S02 and PM07) 
\footnote{The difference from PM07 would come from boundary 
conditions they adopted. 
Since they use an open boundary condition in the symmetry plane 
at the jet basis, this allows the leakage of gas from the boundary 
(see PM07 for more details). Thus, the hot spot pressure drops faster, 
and consequently the growth rate of the hot spot radius is higher. 
These trends for $P_{\rm HS}$ and $r_{\rm HS}$ 
coincide with the deviations between KK06 and PM07. 
Concerning the distinction from S02, the effect of 
relativistic backflow may be a plausible reason 
considering that the deviation appears only for the evolution of 
the hot spot radius. 
If the part of jet power is used to fuel the relativistic 
backflow, its kinetic energy is not available to expand the 
cocoon sideways. As a result, the growth of head and hot spot radius 
would be faster in S02 than KK06. But the deviation (factor 3) 
is not significant even if we consider the long-term evolution of 
radio source, e.g., from $\sim 10^{-3}$ kpc to $10^{3}$ kpc. 
This would imply the relativistic backflow does not have any effect 
on the cocoon dynamics.}. 
However, such small deviations do not affect the following results.

\section{What can we learn from observed $r_{\rm HS}-l_{\rm h}$ relation ?}
In this section, we examine the deceleration and acceleration of hot spot 
velocity by comparing the observed $r_{\rm HS}-l_{\rm h}$ relation 
with the KK06 model. 

\subsection{Constant velocity model or constant aspect ratio model ?}
Here we test two well-known evolution scenarios of radio sources, 
that is, (i) the constant velocity model ($v_{\rm HS}=const$) 
proposed by Readhead et al. (1996a) and 
(ii) the self-similar model with ${\cal R}=const$ (e.g., Falle 1991; 
Begelman 1996; Kaiser \& Alexander 1997).

As seen in Fig. 2, one may think this figure seems to support 
the $v_{\rm HS}=const$ model. However, we should keep in mind 
the large uncertainty of $v_{\rm HS}$ (\S 2.2). 
Furthermore, the deceleration may take place via a strong interaction with 
denser ambient gas in host galaxies, since compact radio sources such as 
CSOs and MSOs would affect a significant interaction with 
the surrounding ambient medium as they propagate through it 
(e.g., Gelderman \& Whittle 1994; de Vries et al. 1997; Axon et al. 2000; 
Gupta et al. 2005; Holt et al. 2008). 
At present, it is still under debate whether $v_{\rm HS}=const$ is really 
correct. In order to test the validity of the $v_{\rm HS}=const.$ model, 
we predict the required mass density profile to explain the observed 
$r_{\rm HS}-l_{\rm h}$ diagram. 
From eq. (2), the relation of $S_{\rm v}\propto 2-X(2-0.5\alpha)=0$ is needed 
for the constant hot spot velocity. 
By eliminating $X$, $\alpha$-dependence of $r_{\rm HS}$ shows 
\begin{equation}
r_{\rm HS}\propto l_{\rm h}^{\alpha/2}. 
\end{equation}

\vspace{5mm}
\epsfxsize=8cm 
\epsfbox{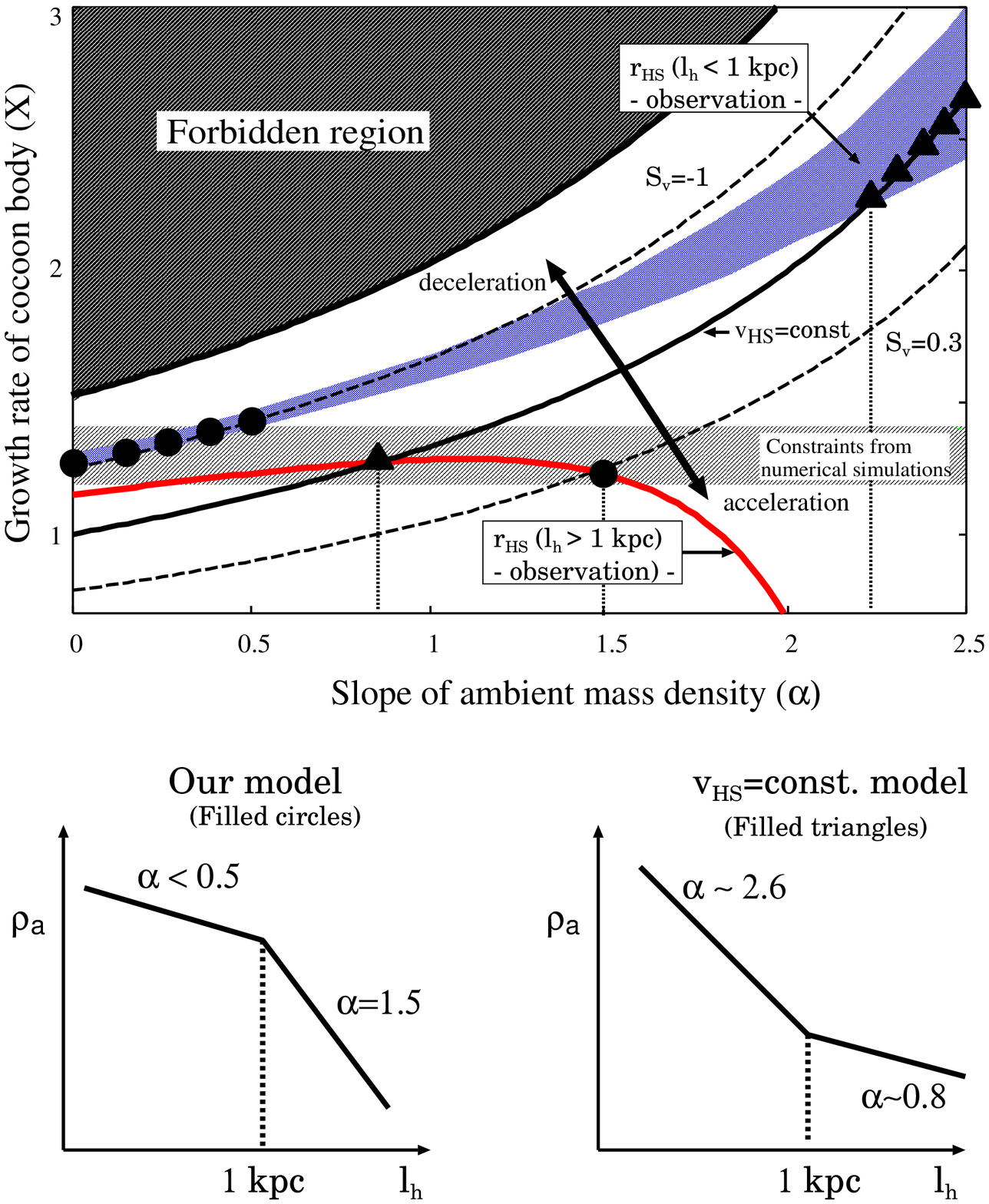}
\figcaption
{
\small Growth rate of cocoon body ($A_{\rm c}\propto t^{X}$) 
against the value of exponents of the ambient mass density profile 
($\rho_{\rm a}\propto l_{\rm h}^{-\alpha}$). 
The blue-shaded region shows the allowed region given by the best fit 
value of the observed $r_{\rm HS}-l_{\rm h}$ relation for the CSO-MSO phase ($l_{\rm h} <$ 1 kpc), i.e., $S_{\rm r}=1.34\pm0.24$. 
The red line corresponds to the allowed region obtained 
for the MSO-FRI\hspace{-.1em}I phase ($l_{\rm h} > $ 1 kpc), 
including the fitting errors, i.e., $S_{\rm r}=0.44\pm0.08$. 
The black solid line shows $v_{\rm HS}$=const ($S_{\rm v}=0$). 
The two black dashed lines are relations for the evolution of 
hot spot velocity $v_{\rm HS}\propto l_{\rm h}^{S_{\rm v}}$. 
The light gray shaded region represents the range of X-parameter 
derived from comparisons between KK06 and numerical simulations. 
The filled circles show the solutions for our model. 
We can constrain on he power-law index $\alpha (l_{\rm h} < 1\,{\rm kpc})$ 
as $0 \leq \alpha < 0.5$. The filled triangles are solutions for 
the $v_{\rm HS}=const.$. The predicted ambient mass density profiles are shown.
}
\vspace{2mm}

Figure 5 displays the allowed region of the growth rate of cocoon body 
($A_{\rm c}\propto t^{X}$) plotted versus $\alpha$ 
(slope of the ambient mass density profile, 
$\rho_{\rm a}\propto l_{\rm h}^{-\alpha}$). 
The blue-shaded region shows the allowed region given by the best fit 
value of the observed $r_{\rm HS}-l_{\rm h}$ relation 
for the CSO-MSO phase ($l_{\rm h} <$ 1 kpc), i.e., $S_{\rm r}=1.34\pm0.24$. 
The red line corresponds to the allowed region obtained 
for the MSO-FRI\hspace{-.1em}I phase ($l_{\rm h} > $ 1 kpc), 
including the fitting errors, i.e., $S_{\rm r}=0.44\pm0.08$. 
The black solid line shows $v_{\rm HS}$=const ($S_{\rm v}=0$). 
The two black dashed lines are relations for the deceleration 
and acceleration of $v_{\rm HS}$, i.e., 
$S_{\rm v}=-1\,\, {\rm and}\,\, 0.3$ (see eq.(2)). 
The light gray shaded region represents the allowed range of $X$-parameter 
derived from comparisons between the KK06 model and relativistic numerical 
simulations (S02 and PM07). 
The dark gray shaded region shows the forbidden region (see $\S 3$). 
As seen in Fig. 5 (filled triangles and also the schematic picture), 
the constant velocity model can reproduce the observed $r_{\rm HS}-l_{\rm h}$ 
relation only when the slope of the inner part of the density profile 
($< 1$ kpc) is steeper than that of the outer part 
($> 1$ kpc), i.e., $\alpha \sim 2.6\,(l_{\rm h} < 1\,{\rm kpc})$ 
and $\alpha \sim 0.8 \,(l_{\rm h} > 1\,{\rm kpc})$. 
However, such a density profile of ambient matter is unrealistic 
because of the slope of the density profile 
in many clusters of galaxies $\alpha \approx 1.5\sim 2$ 
(e.g., Trinchieri et al.1986; Mulochaey \& Zabludoff 1998, 
Mathews \& Brighenti 2003). 
Thus, the $v_{\rm HS}=const$ model can be declined.

Similar to the $v_{\rm HS}=const$ model, it is an open question 
whether the cocoon shape evolves in a self-similar way, 
i.e., ${\cal R}=const$
(De Young 1997; Komissarov \& Falle 1998; O'Dea 1998; Carvalho \& 
O'Dea 2002a, b; PM03). 
Then, we examine whether the ${\cal R}=const$ model can reproduce the 
observed $r_{\rm HS}-l_{\rm h}$ relation. 
As for the self-similar model, the relation of $S_{\rm R}\propto 
X(2.5-0.5\alpha)-3=0$ is needed from eq. (4). 
By eliminating X, $r_{\rm HS}$ can be 
obtained as 
\begin{equation}
r_{\rm HS}\propto l_{\rm h}^{(\alpha+4)/6}.
\end{equation}
Figure 6 is the same as Fig. 5, but the allowed regions for the aspect ratio 
${\cal R}$ are plotted. The black solid line denotes ${\cal R}=const\, 
(S_{\rm R}=0)$ and the other two black solid lines are the solutions of 
$S_{\rm R}=-0.5$ and 0.3. 
As seen in Fig. 6, the self-similar model (${\cal R}=const$) 
cannot explain the observed $r_{\rm HS}-l_{\rm h}$ relation because 
the black solid line does not overlap both blue and red regions. 
Therefore, the case of ${\cal R}=const$ can also be ruled out. 

\vspace{5mm}
\epsfxsize=8cm 
\epsfbox{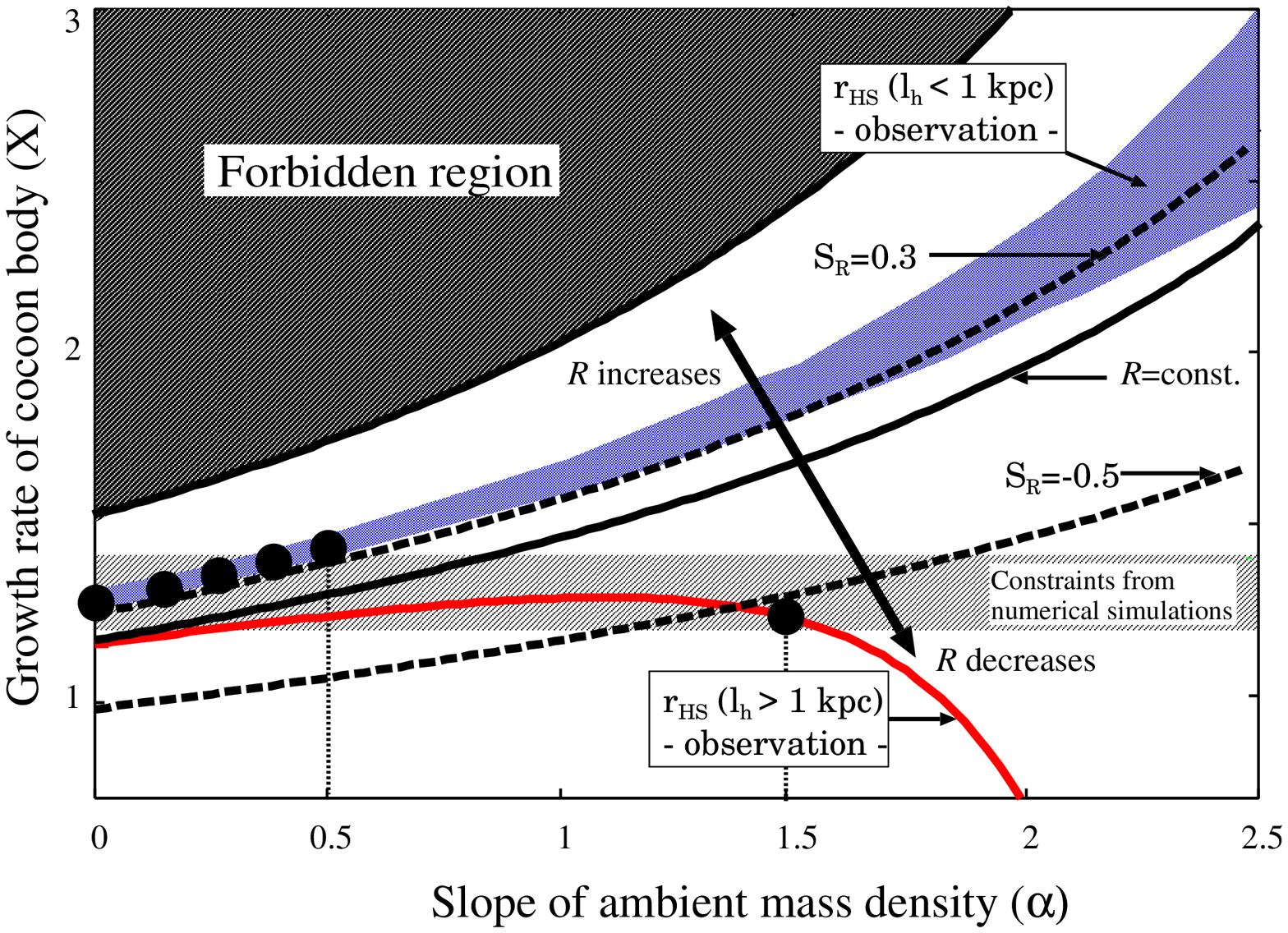}
\figcaption
{
Same as Figure 5, but the black solid line shows 
${\cal R}$=const ($S_{\rm R}=0$) and  
the two black dashed lines are relations of $S_{\rm R}=0.3$ and 0.5 
where ${\cal R} \propto l_{\rm h}^{S_{\rm R}}$.
}
\vspace{2mm}

On the whole, two well-known models are not able to explain the 
observed $r_{\rm HS}-l_{\rm h}$ relation, even if we consider the 
possibility that the actual power-law index is steeper than 
$S_{\rm r} > 1.34\pm 0.24$ for $l_{\rm h} > 1\, {\rm kpc}$ 
(see $\S 2.1$).

\subsection{A new finding: Deceleration and acceleration of hot spots velocity}
Here we show an evolutionary track of radio sources to be consistent with 
the observed evolution of $r_{\rm HS}$ (see Fig. 1). 
In groups of galaxies and clusters of galaxies ($l_{\rm h} > 1\,{\rm kpc}$), 
it is well established that the slope of the density profile 
is $\alpha \approx 1.5-2$ (e.g., Trinchieri et al.1986; Mulochaey \& Zabludoff 1998, 
Mathews \& Brighenti 2003). We assume $\alpha=1.5$ for $l_{\rm h} > 
1\,{\rm kpc}$. 
Comparing the KK06 model with the observed $r_{\rm HS}-l_{\rm h}$ relation 
(Fig. 1), we can determine 
(i) the evolution of the advance speed of hot spots, 
(ii) the slope of mass density distribution for $l_{\rm h} < 1\,{\rm kpc}$ 
and (iii) the evolution of cocoon morphology (see Fig. 5 and Fig. 6).

On the evolution of hot spot velocity, we find that the advance speed of 
the spots and lobes show the deceleration phase (CSO-MSO phase) and 
the acceleration phase (MSO-FRI\hspace{-.1em}I phase) as follows 
(see filled circles showing allowed solutions in Fig. 5): 
In the CSO-MSO phase ($l_{\rm h} < 1\,{\rm kpc}$), $v_{\rm HS}$ decelerates as 
\begin{equation}
v_{\rm HS}\propto l_{\rm h}^{-1}.
\end{equation}
In the MSO-FRI\hspace{-.1em}I phase 
($l_{\rm h} > 1\,{\rm kpc}$), $v_{\rm HS}$ slightly accelerates as 
\begin{equation}
v_{\rm HS}\propto l_{\rm h}^{0.3}.
\end{equation}

With respect to the CSO-MSO phase ($l_{\rm h} < 1\,{\rm kpc}$), 
the flatter density profile ($0 < \alpha \leq 0.5$) is predicted 
in order to satisfy both the observed $r_{\rm HS}-l_{\rm h}$ 
diagram and the constraints from numerical simulations (filled circles 
are allowed solutions in Fig. 5). 
The predicted external mass density 
profile is quite similar to a King-profile distribution, 
i.e., $\rho_{\rm a}\propto [1+(r/r_{\rm c})^{2}]^{0.5\alpha}$, 
as indicated by X-ray observations of elliptical galaxies (Trinchieri et al.
1986). Typical core radii in giant elliptical galaxies are 
$r_{\rm c}=0.5-1\,{\rm kpc}$ (Trinchieri et al. 1986). 
This coincides with the observational evidence since most CSO/MSO hosts 
seem to be elliptical galaxies (e.g., O'Dea 1998) although they show 
the features of distortions and interactions. 
If we consider the mass density distribution we predicted, 
the jets propagate through the denser ambient medium 
during the CSO-MSO phase, compared with the MSO-FRI\hspace{-.1em}I phase. 
Then, the interaction between the jets and the ambient medium is stronger 
in the CSO-MSO phase. 
This leads to the larger velocity of sideways expansion in order to 
maintain the energy conservation in the cocoon. 
Thus, $A_{\rm h}$ (or $r_{\rm HS}$) grows faster in the early phase 
of evolution. 
From the equation of motion along the jet axis (see $\S 3$), 
it is found that the advance speed of hot spots ($v_{\rm HS}$) 
is determined by the linear density of the effective working 
surface, $\rho_{\rm a}A_{\rm h}$, i.e., 
$v_{\rm HS}^{2}\propto (\rho_{\rm a}A_{\rm h})^{-1}$. 
Hence, the deceleration and the acceleration of $v_{\rm HS}$ 
depends on the change of the power-law index of ambient density 
profile $\alpha$ in the MSO phase ($\sim$ kpc).
When $\rho_{\rm a}A_{\rm h}$ increases with $l_{\rm h}$ as 
the CSO-MSO phase ($\rho_{\rm a}A_{\rm h}\propto l_{\rm h}^{2}$), 
the hot spot velocity decelerates.
On the other hand, the advance speed accelerates 
if $\rho_{\rm a}A_{\rm h}$ decreases with $l_{\rm h}$ 
as the MSO-FRI\hspace{-.1em}I phase 
($\rho_{\rm a}A_{\rm h}\propto l_{\rm h}^{-0.6}$). 
In the present work, we impose the idea that the evolution of hot spot velocity is mainly determined by the growth of the effective working surface, 
$A_{\rm h}$. Namely, we did not consider the entrainment of the ambient 
medium in the jets for simplicity (see $\S 3$).  
We should stress that it is a new aspect that $A_{\rm h}$ determines 
the acceleration and deceleration of advance speed of hot spots. 
Note that the study on the entrainment process on jets is 
also important and complementary for 
the deceleration of relativistic jets within $\sim$ 1kpc 
(e.g., Bicknell 1984; Laing 1993: De Young 1993; Komissarov 1994; Laing 1996; 
De Young 1997).

Concerning the evolution of cocoon morphology, we show possible solutions (filled circles in Fig. 6). 
As a result, the evolution of a cocoon shape is predicted as follows:
In the CSO-MSO phase, the shape of a cocoon is nearly spherical as 
\begin{equation}
{\cal R}\propto l_{\rm h}^{0.3}.
\end{equation}
The cocoon shape is elongated as $l_{\rm h}$ increases 
in the MSO-FRI\hspace{-.1em}I phase as 
\begin{equation}
{\cal R}\propto l_{\rm h}^{-0.5}.
\end{equation}

Hence, the aspect ratio of cocoon in the MSO phase can be close to unity, 
compared with that in CSOs and FRI\hspace{-.1em}I sources (see Fig. 7). 
In the CSO-MSO stage, the hot spot decelerates with $l_{\rm h}$, i.e., 
$v_{\rm HS}\propto l_{\rm h}^{-1}$, due to the 
strong interaction with the ambient medium. 
In order to maintain the energy conservation in the cocoon, 
$P_{\rm c}A_{\rm c}v_{\rm HS}=const$, and 
the momentum conservation along the jet axis, 
$P_{\rm HS}A_{\rm h}c=const$, 
the sideways expansion velocity decreases more slowly compared with 
the advance speed of hot spots, 
e.g., $v_{\rm c}\propto l_{\rm h}^{-0.44}$ for 
$v_{\rm HS}\propto l_{\rm h}^{-1}$. 
Thus, the aspect ratio approaches unity since the aspect ratio of the cocoon 
can be described as 
${\cal R}\approx P_{\rm c}/P_{\rm HS}\propto v_{\rm c}^{2}/v_{\rm HS}^{2}$.
In the same manner, it is possible to explain the evolution of the cocoon 
shape in the MSO-FRI\hspace{-.1em}I stage, 
$v_{\rm c}\propto l_{\rm h}^{-0.27}$ for 
$v_{\rm HS}\propto l_{\rm h}^{0.3}$.

\vspace{5mm}
\epsfxsize=8cm 
\epsfbox{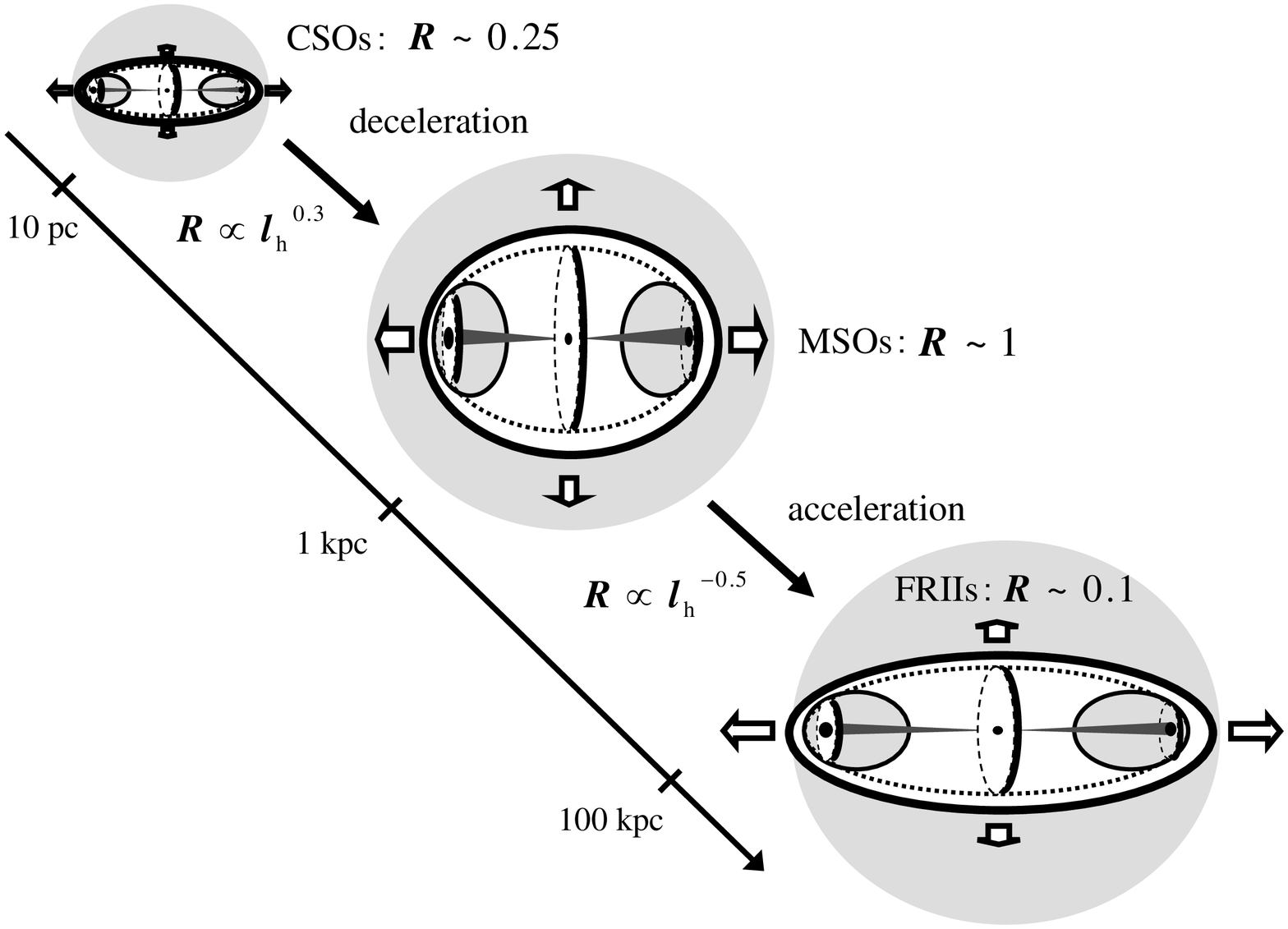}
\figcaption
{
A schematic picture of the evolution of a cocoon morphology 
derived from our predictions (see Fig. 6). 
}
\vspace{2mm}

Recently, we have noticed a very interesting radio source (3C 353) 
whose radio morphology can be well explained by the present model. 
3C 353 is an FRI\hspace{-.1em}I radio galaxy residing on the edge of the cluster and has asymmetric radio lobes, namely the east lobe is spherical 
while the west lobe is elongated. 
Thanks to the new {\it XMM-Newton} observation, it has been found that 
the east lobe is located in the denser region compared with the west lobe 
(Goodger et al. 2008). 
From the above discussions, the spherical structure of the east lobe 
can be understood due to the stronger interaction with the ambient medium 
than the west lobe. 
If this is the case, it is predicted that the advance speed of the 
east lobe and east hot spot is slower than that of the west lobe. 
In future, the advance speed of the east lobe and hot spot will approach 
the sound velocity of the ambient medium. 
Then, the east lobe's type of radio morphology will change from 
FRI\hspace{-.1em}I into FRI type. Such radio galaxies would be 
observed as hybrid morphology radio sources (HYMORS), which have clear 
radio lobes with two different morphologies (e.g., Gopal-Krishna \& Wiita 2000; Gawro\'{n}ski et al. 2006). In order to understand the origin of HYMORS, i.e., the 
origin of FRI/FRI\hspace{-.1em}I dichotomy, it would be important to measure 
the mass density profile around HYMORS. 

\section{Prediction of fate of CSOs: Dead or Alive ?} 
In $\S 4$, we found that the observed 
$r_{\rm HS}-l_{\rm h}$ relation indicates that 
the advance speed decelerates in the CSO-MSO stage and 
accelerates in the MSO-FRI\hspace{-.1em}I stage. 
However, some CSOs would evolve into some other type of radio source 
like low-power extended radio galaxies (FRI radio galaxies), 
since $\leq$ 10 kpc scale dying MSOs, which do not evolve into 
FRI\hspace{-.1em}Is, have been discovered (Parma et al. 2007). 
Thus, we need to further investigate in this section 
the fate of CSOs, or which kinds of CSOs can evolve 
into FRI\hspace{-.1em}I sources. 
To this aim, we compare the predicted evolution of $v_{\rm HS}$ with 
the sound velocity of the ambient medium, $c_{\rm s}$, 
because the cocoon can expand only when $v_{\rm HS} > c_{\rm s}$.
As for the slope of ambient matter density, 
we assume $\alpha \,(l_{\rm h} < 1\,{\rm kpc}) =0$ and 
$\alpha \,(l_{\rm h} > 1\,{\rm kpc}$)=1.5 (see $\S 4.2$). 
Correspondingly, the behavior of $v_{\rm HS}$ can be determined as
$v_{\rm HS}\propto l_{\rm h}^{-1}$ for $l_{\rm h} <\, 1{\rm \,kpc}$ 
and $v_{\rm HS}\propto l_{\rm h}^{0.3}$ for $l_{\rm h} >\, 1{\rm \,kpc}$ 
(see eq. (7) and (8)). 
The hot ambient-gas temperature, $T_{\rm g}$ is  measured 
in the range of $T_{\rm g}=5\times 10^{6}\,{\rm K}-2\times 10^{7}\,{\rm K}$, 
i.e., $c_{\rm s}=(5kT_{\rm g}/3m_{\rm p})^{1/2}
\approx 7\times10^{-4}c-1.4\times10^{-3}c$, 
where $c_{\rm s}$ is the sound velocity of the ambient medium 
(e.g., Trinchieri et al. 1986), $k$ is the Boltzman constant and 
$m_{\rm p}$ is the proton mass. 

Figure 8 shows the evolution of hot spot velocity for three initial 
advance speeds of the ambient medium with 
$v_{\rm HS}(l_{\rm h, 2D})=0.01c, 0.1c\, {\rm and}\, 0.5c$, 
where $l_{\rm h, 2D}\equiv v_{\rm HS}(l_{\rm h, 2D})t_{\rm min}$ 
is the distance from the core at which the 2-D phase 
($A_{\rm h}$ growth phase) starts. 
Here we suppose $l_{\rm h, 2D}=5\,{\rm pc}$ from Fig. 1. 

\vspace{5mm}
\epsfxsize=8cm 
\epsfbox{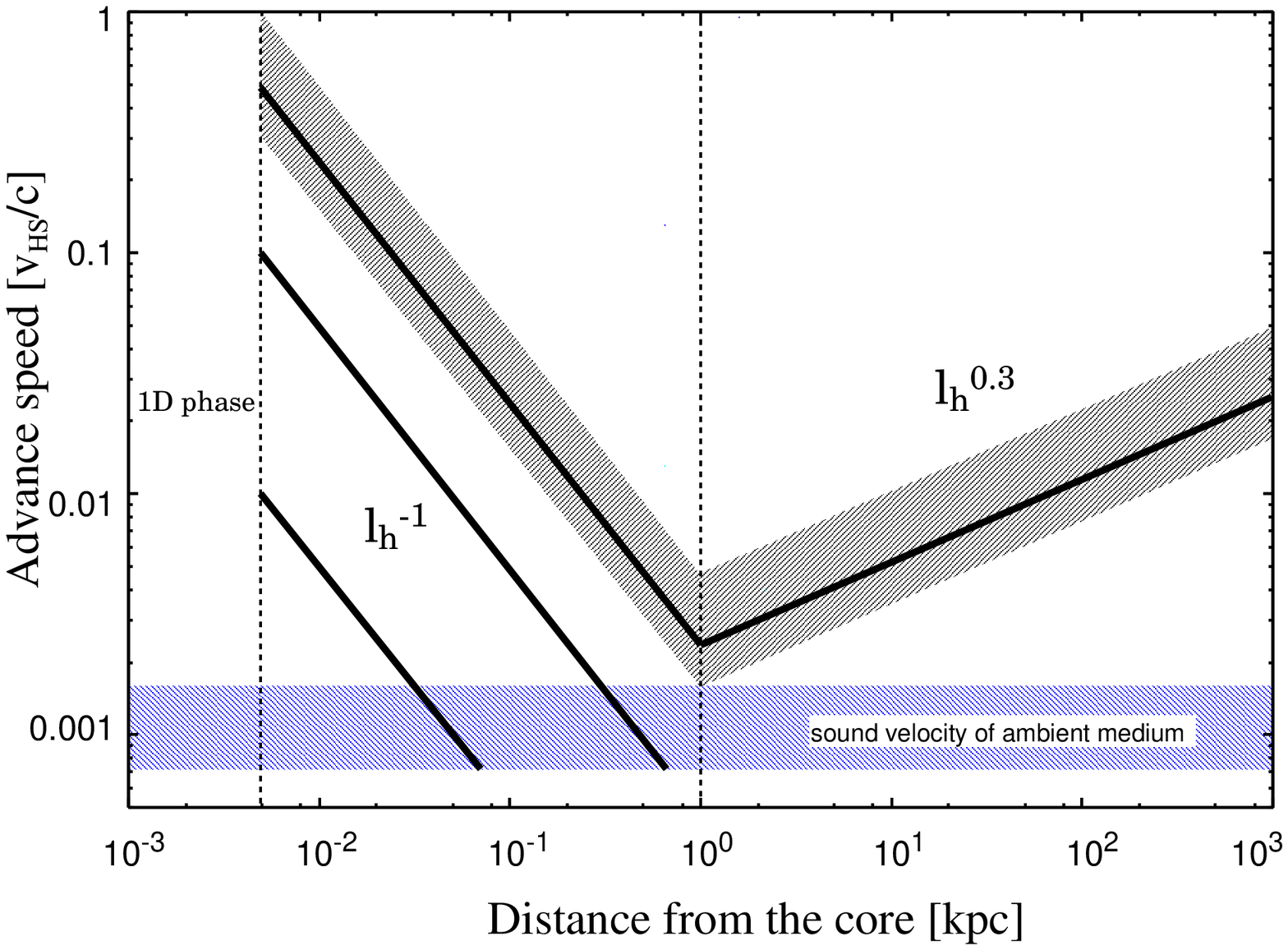}
\figcaption
{
Our predictions of evolution of $v_{\rm HS}$, i.e., 
$v_{\rm HS}\propto l_{\rm h}^{-1}$ for $l_{\rm h} <\, 1{\rm \,kpc}$ 
and $v_{\rm HS}\propto l_{\rm h}^{0.3}$ for $l_{\rm h} >\, 1{\rm \,kpc}$. 
The black solid lines denote the evolution of $v_{\rm HS}$ 
for $v_{\rm HS}(l_{\rm h,2D})=0.01\, c, 0.1\, c,\,{\rm and}\,0.5\, c$ 
where $l_{\rm h, 2D}=5\, {\rm pc}$. 
The black-shaded region represents the evolutionary path from 
CSOs into FRI\hspace{-.1em}Is.
The blue-shaded region shows the range of sound velocity of the 
ambient medium, i.e., $7\times 10^{-4}c < c_{\rm s} < 1.4\times 10^{-3} c$ 
($5\times 10^{6}\, {\rm K} < T_{\rm g} < 2\times 10^{7}\, {\rm K}$) 
where $T_{\rm g}$ is the temperature of the ambient medium.
}
\vspace{2mm}

As seen in Fig. 8, we find that CSOs can evolve into FRI\hspace{-.1em}I 
sources, passing through MSOs for any $l_{\rm h}$ when $v_{\rm HS}(l_{\rm h, 2D})$ is larger than about 0.1c because of $v_{\rm HS}(l_{\rm h}) > c_{\rm s}$. 
On the other hand, when $v_{\rm HS}(l_{\rm h, 2D})$ is less than about 0.1c, 
$v_{\rm HS}$ is comparable to the sound velocity of hot spots 
during the CSO-MSO phase ($l_{\rm h} < 1\, {\rm kpc}$). 
When $v_{\rm HS}$ equals to $c_{\rm s}$, the matter and energy of the cocoon 
mix with the ambient medium. After that, the cocoon expands subsonically 
like a sonic boom (Churazov et al. 2000, 2001; Reynolds et al. 2001; 
Brighenti \& Mathews 2002; Zanni et al. 2003). 
These results indicate that the advance speed of hot spots 
during the 1 D phase (constant $A_{\rm h}$) controls the fate of CSOs. 
Thus, the fate of CSOs can be divided into two cases as follows:
\begin{enumerate}
\item The fate of CSO with low $v_{\rm HS}(l_{\rm h, 2D})$ $< \sim0.1\, c$:\\
The hot spots velocity is comparable to the sound velocity of ambient matter 
at the MSO stage ($< 1\, {\rm kpc}$). 
Consequently, the ambient gas may be heated up since the relativistic plasma 
within the cocoon leaks into the ambient medium. The cocoon morphology is 
distorted in the MSO phase. After that, radio sources would 
expand with the sound velocity. 
Thus, the evolutionary sequence appears as follows: 
\begin{center}
CSO $\to$ MSO $\to$ FRI.
\end{center}

\item The fate of CSO with high $v_{\rm HS}(l_{\rm h, 2D})$ $> \sim0.1\, c$:\\
The CSOs can evolve into FRI\hspace{-.1em}I sources, passing through MSOs whose cocoon shape is spherical. 
Namely, the evolutionary sequence is as follows: \begin{center}
CSO $\to$ MSO $\to$ FRI\hspace{-.1em}I.
\end{center}
\end{enumerate}

The above predicted cocoon morphology of MSOs may coincide with observations 
in which the fraction of distorted morphology of MSOs is higher than that of CSOs (Saikia et al. 1995; O'Dea 1998; Dallacasa et al. 2002a, b; 
Parma et al. 2007). 
Moreover, with respect to case (1), Drake et al. (2004) observed 
radio-weak CSOs from the IRAS survey, then they found that radio-weak CSOs 
are young ($< 10^{6}\,{\rm yr}$) and show slightly irregular morphology radio 
sources (also see Kunert-Bajraszewska et al. 2005). 
Therefore, such radio-weak CSOs may be young counterparts of 
$\sim 1\,{\rm kpc}$ scale low-power compact radio sources 
(Giroletti et al. 2007).
Such low-power CSOs might be observed as the dying MSOs 
(e.g., Parma et al. 2007).  
As another possibility, low-power CSOs may be the progenitors 
of FRIs because the radio lobes expand with $\approx c_{\rm s}$, and then 
the projected linear size of radio sources can reach 
$\sim 30\,{\rm kpc}\,(=10^{-3}c\times 10^{8}\,{\rm yr})$ 
if the typical age of FRIs is $\approx 10^{8}\,{\rm yr}$ 
(Parma et al 1999). 
If this is the case, the origin of the FRI/FRI\hspace{-.1em}I dichotomy could 
be related to the difference in initial advance speed of hot spots $v_{\rm HS,1D}$ which depend on $L_{\rm j}$ and $\rho_{\rm a}$ (see $\S 3$).

Moreover, these findings impact on AGN feedback associated 
with AGN jets. The influence of AGN activity on the evolution of 
the host galaxy is now a hot and timely topic. 
It has been suggested that the AGN bubbles quench 
star formation and regulate the SMBH growth, 
and consequently the local BH-to-bulge relation can be well reproduced 
(e.g., Silk \& Rees 1998; King 2003). 
However, we should recall that their conclusions rely on the dynamical 
evolution of AGN bubbles with the constant shock velocity ($v_{\rm HS}=const$), 
assuming a singular isothermal hot-ambient distribution ($\rho_{\rm a}\propto r^{-2}$). However, as we mentioned the constant velocity model 
does not match with our predictions derived from 
the actual evolution of young radio sources. Therefore, we must reconsider 
the conditions that AGN feedback due to AGN jets works effectively. 
This is left for our future work.

Finally it is of value to mention what we should do as future observations, 
to reveal the dynamical evolution of radio sources as follows:

\begin{enumerate}
\item
We need to evaluate the accurate $v_{\rm HS}$ of MSOs 
in future VLBI observations, to test the deceleration and acceleration of 
hot spot velocity. 
At the same time, it is important to estimate the advance velocity of 
hot spots for FRI\hspace{-.1em}Is by kinematic studies 
because most $v_{\rm HS}$ for FRI\hspace{-.1em}Is have been 
derived by synchrotron-aging work only. 
\item
In order to confirm the evolution of the cocoon shape we predict here 
(see Fig. 7), it is worth systematically measuring the aspect ratio of cocoons for CSOs, MSOs and FRI\hspace{-.1em}Is. 
But the emission of the cocoon body is very faint in general due to the 
synchrotron cooling, and thus it would be hard to assess the clear shape of 
a cocoon (i.e., $A_{\rm c}$) by the present facility. 
The sensitivity and low frequency capability of SKA can be used for 
direct measurement of cocoon morphology, since the cooling effect 
is smaller at lower frequency. 
Instead of the direct measurement of the cocoon shape, it is also important to 
measure the morphology of radio lobes, i.e., $A_{\rm h}^{1/2}/l_{\rm h}$ 
(see Fig. 4) as far as $A_{\rm c}\propto A_{\rm h}$ is a reasonable assumption. \item
In the present paper, we impose the idea that the initial advance velocity of hot spots, $v_{\rm HS}(l_{\rm h, 2D})$, may be interpreted by the differences in 
total kinetic energy. 
However, we cannot rule out the possibility that 
the ambient mass density controls $v_{\rm HS}(l_{\rm h, 2D})$. 
Thus, it is crucial to systematically explore whether there is a significant difference in ambient mass density for CSOs, MSOs and FRI\hspace{-.1em}Is. 
For this purpose, the search for cold gas via molecular gas and 21 cm HI is 
very useful using future facilities with high spatial resolution such as 
ALMA and SKA. At the same time, the study for diffuse hot gas will be 
also important using future X-ray satellites with high sensitivity and 
high spatial resolution such as XUES and Constellation-X.
\item
In order to reveal the end of the 1D phase of hot spot evolutions, 
it is necessary to discover and elucidate the candidates for ultra compact 
symmetric radio sources (e.g., $l_{\rm h} < 5\,{\rm pc}$) 
such as High Frequency Peakers (HFPs) that have been 
intensively observed and a growing number of the sources are being detected 
(e.g., Dallacasa et al. 2000; Tinti et al. 2005; Orienti et al. 2007), 
using high resolution VLBI such as VSOP-2 (see also Snellen 2008). 
This is very important not only to predict the fate of CSOs as we mentioned, 
but also to reveal the physical conditions for the transitions from 1D phase 
to 2D phase. 
\item 
It is worth exploring whether radio-weak CSOs, including radio-quiet AGNs 
(Seyferts and radio-quiet quasars) follow the same $r_{\rm HS}-l_{\rm h}$ 
diagram. The present model predicts that the observed $r_{\rm HS}-l_{\rm h}$ 
relation of bright radio sources holds radio-weak sources 
as far as the AGN jets are composed of the relativistic plasma. 
So far, it has been very difficult to measure 
the hot spot properties (e.g., size and velocity) because of 
the faintness of low-powered jet sources. However, future 
low-frequency radio telescopes (e.g., LOFAR and SKA) are ideal tools 
for understanding the dynamical evolution of radio-weak sources 
from the point of view of sensitivity and the low-frequency capability.  
\end{enumerate}

\section{Summary}
We have investigated the physical relationship between CSOs, 
MSOs, and FRI\hspace{-.1em}Is, together with the coevolution model 
of hot spots and a cocoon (KK06) and the observation reflecting the 
dynamical evolution of radio sources. For a larger sample of radio sources 
than previous work (J00; PM03), we find that the evolution of the hot spot radius clearly changes at $l_{\rm h}\sim$ 1kpc, even after we carefully exclude the observational bias. Based on these observational characteristics of hot spots, 
we investigate the evolutionary track of extragalactic radio galaxies. 
The main results of the present work are summarized as follows: 

(i) We test two well-known evolution models of radio sources, i.e., 
``the constant velocity'' model and ``the constant aspect ratio'' model. 
Since the mass density profile of the ambient matter shows unrealistic 
behavior, i.e., the derived power law index ($\alpha\approx 0.8$) is 
shallower than $\alpha=1.5-2$ in groups of galaxies and clusters of galaxies, 
the constant velocity model can be ruled out. 
In the case of the self-similar model, it turns out that 
the constant aspect ratio model cannot reproduce the observed 
$r_{\rm HS}-l_{\rm h}$ relation. 
Therefore, the case of constant aspect ratio can also be declined.

(ii) We investigate the evolution of hot spot velocity to reproduce 
the observed $r_{\rm HS}-l_{\rm h}$ relation. As a result, 
we find that the advance speed of hot spots and lobes 
strongly decelerate when the jets pass through the ambient medium in 
host galaxies (i.e., the CSO-MSO phase), while the jets slightly accelerate 
outside host galaxies (i.e, the MSO-FRI\hspace{-.1em}I phase). 
Recent observation shows that a possible correlation between 
the outflow velocity of the ionized gas and $l_{\rm h}$ 
might indicate a possible deceleration of the jets as 
they pass through the host galaxy (Labiano 2008). 
This trend seems to be consistent with the preferred model, 
although the available data is not enough to compare with our prediction. 
Furthermore, we can constrain the allowed value of $\alpha$ as 
$0 \le \alpha <0.5$ for $l_{\rm h} < 1\,{\rm kpc}$. 
The main origin of the deceleration and acceleration of hot spots is 
the change in power-law index of ambient density distribution 
at $\sim$ 1 kpc. It is worth emphasizing again that the advance speed 
of hot spots is determined by the growth of the cocoon head, which is 
a new aspect of the deceleration and acceleration of relativistic jets.

(iii) It is also found that the cocoon shape of MSOs becomes spherical 
or disrupted, while the cocoon morphology of CSOs and FRI\hspace{-.1em}Is is 
elongated. The radio image of MSOs seems to be consistent with our prediction 
(Dallacasa et al. 2002a, b). However, the emission of the cocoon is generally 
very faint due to the synchrotron cooling and thus it is quite difficult to 
observe it using present facilities. Thus, it is of value to systematically 
measure the aspect ratio (or the morphology of radio lobes) for CSOs, MSOs, and FRI\hspace{-.1em}Is using future low-frequency radio telescopes (e.g., LOFAR 
and SKA).

(iv) We predict the fate of CSOs, comparing the advance speed of hot spots with the sound velocity of the ambient medium. Only CSOs whose initial advance speed is higher than about 0.1c can be progenitors of 
powerful FRI\hspace{-.1em}Is. It would be in agreement with the observation 
which the fraction of FRI\hspace{-.1em}Is is less than that of CSOs. 
Moreover, the fate of CSOs would be closely related to the 
FRI/FRI\hspace{-.1em}I dichotomy and AGN feedback which we will discuss 
in our up-coming paper.

\acknowledgments 
We would like to thank A. Cellotti, I. A. G. Snellen, L. Danese, 
M. Nakamura, S. Inoue, S. Jeyakumar and T. Nagao for fruitful discussions. 
We also thank an anonymous referee for useful comments. 
NK is financially supported by the Japan Society for the Promotion of Science 
(JSPS) through the JSPS Research Fellowship for Young Scientists. 
MK acknowledges the Grant-in-Aid for 
Scientific Research of the Japanese Ministry of Education, Culture, Sports, 
Science and Technology, No. 14079025. 
This research has made use of the NASA/IPAC Extragalactic Database
(NED) which is operated by the Jet Propulsion Laboratory, California
Institute of Technology, under contract with the National Aeronautics
and Space Administration.

\newpage
\begin{table}
  \caption{Physical parameters of hot spots}
\begin{center}
see http://yso.mtk.nao.ac.jp/$^{\sim}$kawakatu/Kawakatu08.pdf
\end{center}
\end{table}

\newpage
\begin{table}
  \caption{Advance speed of Hot spots}
\begin{center}
see http://yso.mtk.nao.ac.jp/$^{\sim}$kawakatu/Kawakatu08.pdf
\end{center}
\end{table}

\newpage
\begin{table}
  \caption{Apparent separation rate ($v_{\rm app}$) between hot spots}
  \label{Vapp_table}
  \begin{center}
    \begin{tabular}{cccc} \tableline\tableline
Source & z & $v_{\rm app}$ [$\mu$as]  &	Reference	\\ \tableline 
0108+388 & 0.699 & 9.27$\pm$1.21 & (1)		\\
0710+439 & 0.518 & 17$\pm$1 & (2)		\\
1031+567 & 0.4597 & 14.6$\pm$4.8 & (3)		\\
1245+676 & 0.1071 & 34.9$\pm$1.8 & (4)			\\
1943+546 & 0.263 & 25$\pm$2 & (2)			\\
2352+495 & 0.238 & 32.7$\pm$11 & (3)			\\
OQ208 &	0.0766 & 32$\pm$20 & (5)			\\
CTD93 &	0.473 &	23$\pm$7.7 & (6)				\\
3C84\tablenotemark{a}	&	0.0183	&	270	$\pm$	70	&	(7)		\\
							
J1111+1955	&	0.299	&	$<10$			&	(8)			\\
J1414+4554	&	0.19	&	$<14$			&	(8)			\\
J1734+0926	&	0.61	&	$<8$			&	(8)			\\
1946+708	&	0.101	&	$<60$			&	(9)		\\
J1934-638\tablenotemark{a}	&	0.183	&	$<30$			&	(10)		\\ \tableline 
\end{tabular}
\tablenotetext{a}{$v_{\rm app}$ is measured between the core and hot spot.}
\tablerefs{(1) Owsianik, Conway \& Polatidis 1998, (2) Polatidis \& Conway 2003, (3) Taylor et al. 2000, (4) Marecki et al. 2003, (5) Wang et al. 2003, 
(6) Nagai et al. 2006, (7) Asada et al. 2006, (8) Gugliucci et al. 2005,
 (9) Taylor \& Vermeulen 1997, (10) Tzioumis et al. 1989
}
\end{center}
\end{table}

\begin{table}[htb]
  \caption{Comparison with 2D relativistic hydrodynamic simulations}
  \label{Comp_num}
\begin{center}
{
\begin{tabular}{cccc}
\hline \hline
----- &  $v_{\rm HS}$ & $r_{\rm HS}$ & $P_{\rm HS}$ \\
\hline
 & & (i) A uniform ambient medium ($\alpha=0$)$^{a}$ & \\
S02 & $l_{\rm h}^{-0.55}$ & $l_{\rm h}^{0.45}$ & $l_{\rm h}^{-1.1}$ \\
KK06 & $l_{\rm h}^{-0.56}$ & $l_{\rm h}^{0.55}$ & $l_{\rm h}^{-1.1}$ \\ 
\hline
 & & (ii) A declining ambient medium ($\alpha=1$)$^{a}$ & \\
PM07 & $l_{\rm h}^{-0.11}$ & $l_{\rm h}^{0.78}$ & $l_{\rm h}^{-1.4}$ \\
KK06 & $l_{\rm h}^{-0.11}$ & $l_{\rm h}^{0.67}$ & $l_{\rm h}^{-1.2}$ \\\hline
\end{tabular}
}
\noindent
\end{center}
{NOTE.--
$^{a}$The case (i) corresponds to KK06 model with $X=1.2$, 
while the case (ii) corresponds to KK06 model with $X=1.4$}
\end{table}

\end{document}